\documentclass[pre,aps,10pt,superscriptaddress,twocolumn,floatfix]{revtex4-2}

\usepackage[nice]{nicefrac}
\usepackage{graphicx,color}
\usepackage{amsmath,amssymb,bm,bbm}
\usepackage{amsmath}
\usepackage{braket}
\usepackage{float}
\usepackage{placeins}
\usepackage{graphicx}
\usepackage[T1]{fontenc}

\usepackage[plainpages=false,pdfpagelabels,colorlinks=true,linkcolor=red,urlcolor=blue,citecolor=blue,pdftitle={},pdfauthor={},pdfdisplaydoctitle=true,pdfduplex=DuplexFlipLongEdge]{hyperref}

\definecolor{darkred}{rgb}{0.90,0.2,0.2}
\definecolor{darkgreen}{rgb}{0,0.60,.2}
\definecolor{darkblue}{rgb}{0.1,0.3,1}
\definecolor{grey}{cmyk}{0,0,0,0.25}
\definecolor{orange}{cmyk}{0,0.6,0.8,0}

\begin{document}
\title{Localization transitions in quadratic systems without quantum chaos}

\author{Mateusz Lisiecki}
\affiliation{Institute of Theoretical Physics, Wroclaw University of Science and Technology, 50-370 Wroc{\l}aw, Poland}
\author{Lev Vidmar}
\affiliation{Department of Theoretical Physics, J. Stefan Institute, SI-1000 Ljubljana, Slovenia}
\affiliation{Department of Physics, Faculty of Mathematics and Physics, University of Ljubljana, SI-1000 Ljubljana, Slovenia\looseness=-1}
\author{Patrycja  \L yd\.{z}ba}
\affiliation{Institute of Theoretical Physics, Wroclaw University of Science and Technology, 50-370 Wroc{\l}aw, Poland}

\begin{abstract}
Transitions from delocalized to localized eigenstates have been extensively studied in both quadratic and interacting models.
The delocalized regime typically exhibits diffusion and quantum chaos, and its properties comply with the random matrix theory (RMT) predictions.
However, it is also known that in certain quadratic models, the delocalization in position space is not accompanied by single-particle quantum chaos.
Here, we study the one-dimensional Anderson and Wannier-Stark models that exhibit eigenstate transitions from localization in quasimomentum space (supporting ballistic transport) to localization in position space (with no transport) in a nonstandard thermodynamic limit, which assumes rescaling the model parameters with the system size.
We show that the transition point may exhibit an unconventional character of Janus type, i.e., some measures hint at the RMT-like universality, while others depart from it.
For example, the eigenstate entanglement entropies may exhibit, depending on the bipartition, a volume-law behavior that aligns with that of Haar-random Gaussian states, or its coefficient converges to a lower, nonuniversal value.
Our results hint at a rich diversity of volume-law eigenstate entanglement entropies in quadratic systems that are not maximally entangled.
\end{abstract}
\maketitle

\section{Introduction}

The concept of quantum chaos in lattice systems represents a remarkable phenomenon that helps establish connections between different fields of physics, e.g., between ergodicity and thermalization in statistical physics and diffusive transport in condensed matter physics~\cite{dalessio_kafri_16,Bertini_heidrichmeisner_21,sierant_lewenstein_24,PhysRevE.60.3949}.
Focusing on quadratic systems, the prototypical manifestation of single-particle quantum chaos is the appearance of pseudo-random amplitudes of one-body Hamiltonian eigenfunctions when written in a non-fine-tuned basis.
Consequently, chaotic behavior can be characterized using various measures, e.g., the spectral properties complying with the random-matrix theory (RMT) predictions~\cite{PhysRevLett.52.1,PhysRevLett.77.4744,Ul_akar_2022} or the entanglement entropy of typical Hamiltonian eigenstates complying with the corresponding Haar-random pure states~\cite{PRXQuantum.3.030201,PhysRevLett.125.180604,PhysRevB.103.L241118,Lydzba_2021}.
Moreover, the breakdown of quantum chaos also serves as a signature of the opposite behavior, e.g., the emergence of localization.
A paradigmatic example is the single-particle localization in the 3D Anderson model, where the breakdown of the agreement between the single-particle spectrum and the RMT prediction has been used for over 30 years as a method to pinpoint the localization transition.~\cite{altshuler_zharekeshev_88,PhysRevB.47.11487,Braun_1995,Suntajs_2021}.

Nevertheless, not all physical systems exhibit quantum chaos.
Notable counter-examples are quasiperiodic quadratic models, with the Aubry–André model being a well-known example~\cite{Harper_1955,Aubry1979}.
In the latter, the Hamiltonian eigenstates are localized in quasimomentum space at weak quasiperiodic potentials and localized in position space at strong quasiperiodic potentials~\cite{PhysRevB.40.8225,PhysRevB.34.2041,Suslov2008}.
To emphasize that none of these regimes exhibits quantum chaos, we refer to this eigenstate transition as quasimomentum-to-position space localization transition, or localization-to-localization transition for short.
Despite the absence of the single-particle quantum chaos reference point, the transition point is known using analytical arguments~\cite{Aubry1979} and the numerical measures such as the inverse participation ratio of the wave function coefficients are consistent with these predictions~\cite{PhysRevB.100.195143,Dom_nguez_Castro_2019}.

Here, we study other quadratic models of spinless fermions in one-dimensional lattices with $L$ sites that exhibit the localization-to-localization transition in the absence of single-particle quantum chaos.
We focus on the tight-binding chain subjected to random disorder with the amplitude $W$ (i.e., the 1D Anderson model) and compare results to the case where the disorder is replaced by a linear potential of strength $F$ (i.e., the 1D Wannier-Stark model).
In the thermodynamic limit, the single-particle eigenstates are localized in quasimomentum space for vanishing on-site potentials (they are plane waves under periodic boundary conditions), while they are localized in position space for any nonzero potential~\cite{Thouless1983,Thouless1979,Thouless1974,Kappus1981,Tessieri_2012,PhysRevB.8.5579,van_Nieuwenburg_2019}.
However, in the so-called nonstandard thermodynamic limit, where we rescale the on-site potentials with the system size as $W \to \tilde{W} = W\sqrt{L}$ and $F \to \tilde{F} = F L$, both models exhibit localization-to-localization transitions at nonzero $\tilde{W}$ or $\tilde{F}$. We note that the concept of nonstandard thermodynamic limit has been previously introduced in single-particle~\cite{PhysRevE.100.022142,PhysRevE.86.011142} and many-body~\cite{PhysRevB.97.214205,Gopalakrishnan_2019,PhysRevB.109.214203,PhysRevB.108.174303,PhysRevB.105.214308} systems,  although the specific form of rescaling may differ.

We study various transition indicators and we ask to what degree does the the localization-to-localization transition point exhibit special (possibly universal) properties.
For the 1D Anderson model, we find that the localization-to-localization transition point has a Janus-type character, i.e., certain properties display an emergent agreement with the RMT universality, while others depart from it. This is, to our knowledge, the first example of such dual behavior, and the full mechanism behind this phenomenon remains to be understood.

\begin{figure}[t!]
\includegraphics[width=\columnwidth]{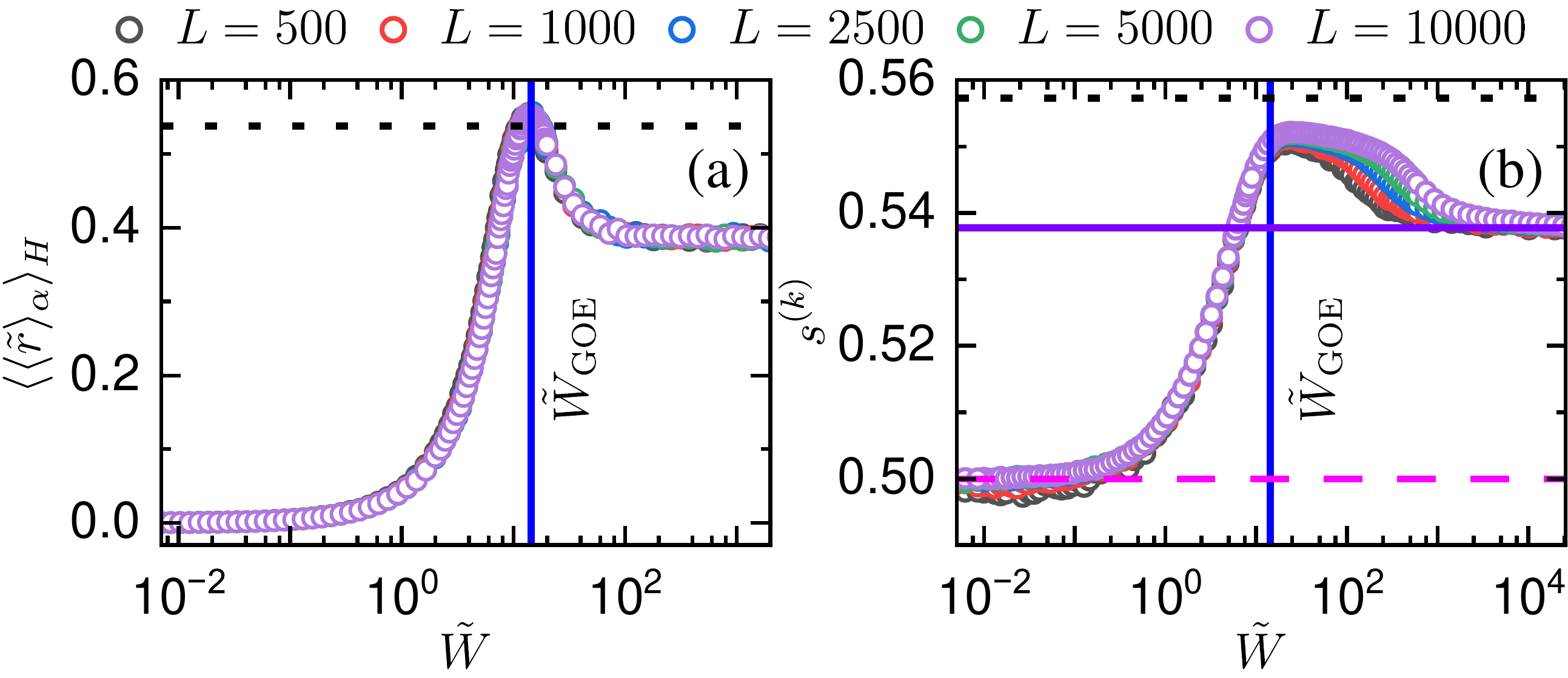}
\caption{(a) The level spacing ratio, introduced in Sec.~\ref{sec_ss}, and (b) the volume law coefficient $s^{(k)}$ of the entanglement entropy in the quasimomentum space, introduced in Sec.~\ref{sec:bipartitions_momentum}, plotted against the scaled disorder $\tilde{W}=W\sqrt{L}$ in the Anderson model with periodic boundary conditions. We consider system sizes $L=500, 1000, 2500, 5000$ and $10000$. The blue solid vertical lines in (a) and (b) mark $\tilde{W}_\text{GOE}$, which is established from the level spacing ratio of the Anderson model with open boundary conditions, see App.~\ref{Appendix_spectral}, and it accurately matches with the position of the peak in (a). The violet solid horizontal line in (b) marks the free fermion results, $s^{(i)}_\text{free}$~\cite{Vidmar_2017}. The black dotted horizontal lines in (a) and (b) mark the predictions for quantum-chaotic quadratic Hamiltonians $\tilde{r}_\text{GOE}$~\cite{PhysRevLett.110.084101} and $s^{(i)}_\text{chaotic}$~\cite{PhysRevLett.125.180604}, respectively. Numerical results in (a) were averaged over $100$ single-particle energy eigenstates in the middle of the spectrum and $100$ disorder realizations, while in (b) they were averaged over $100$ many-body energy eigenstates in the middle of the spectrum and $500$ disorder realizations. Legend is provided above the figure.}
\label{fig_1}
\end{figure}

We first show that the detection of the localization-to-localization transition point appears to be associated with a higher degree of ambiguity than the detection of chaos-to-localization transition.
The latter is commonly studied in the literature and it is often characterized by a scale-invariant point of some simple chaos indicators, such as the level spacing ratio or the eigenstate entanglement entropy~\cite{PhysRevB.107.064205,PhysRevB.102.064207,Abanin_2019,De_2021,PhysRevB.99.104205}.
In contrast, for the localization-to-localization transitions studied here, due to the absence of a quantum chaos reference point, it is not always clear which indicators can detect the transition.
Nevertheless, some measures, such as those shown in Fig.~\ref{fig_1} for the 1D Anderson model, reveal a clear distinction between the two localized regimes, hinting at the emergence of a special point at the boundary between them.

After determining the position of the transition point, we study its properties, focusing on the bipartite eigenstate entanglement entropies of many-body Hamiltonian eigenstates.
We observe that the average eigenstate entanglement entropy for a bipartition in quasimomentum space [shown in Fig.~\ref{fig_1}(b)] approaches the value of Haar-random Gaussian states~\cite{PhysRevLett.125.180604,PhysRevB.103.L241118} at the transition, hinting at the RMT universality.
In contrast, the average eigenstate entanglement entropy for a bipartition in position space [studied in Sec.~\ref{sec:bipartitions_real}] appears to have no connection to the value expected for Haar-random Gaussian states and can also differ from the value observed for translationally invariant free fermions~\cite{Vidmar_2017}.
These results hint at a rich diversity of volume-law entanglement entropies in quadratic (but also integrable interacting) models and reveal a nontrivial structure of Hamiltonian eigenstates at the localization-to-localization transitions.

The paper is organized as follows.
In Sec.~\ref{sec:models} we introduce the models and in Sec.~\ref{sec:nonstandard} we discuss the emergence of localization-to-localization transition in the nonstandard thermodynamic limit, focusing on the time evolution of wave packets and inverse participation ratios.
We then introduce the level spacing measures of the single-particle spectrum in Sec.~\ref{sec_ss} and the entanglement entropy measures of many-body eigenstates in Sec.~\ref{sec_ee}.
In the latter, we consider the bipartitions in both position and quasimomentum spaces.
We conclude in Sec.~\ref{sec:conclusions}.

\section{Quadratic fermionic models} \label{sec:models}

We consider two quadratic models on a one-dimensional (1D) lattice with $L$ sites that preserve the particle number. 
The first model is the Anderson model,
\begin{equation}
     \hat{H}_\text{A}=-t\sum_{i} \left( \hat{c}_i^\dagger \hat{c}_{i+1}+\text{h.c.}\right)+\frac{W}{2}\sum_{i} r_i \hat{c}_i^\dagger \hat{c}_{i},
     \label{eq_ha}
\end{equation}
where $\hat{c}_i^\dagger$ ($\hat{c}_i$) creates (annihilates) a spinless fermion at site $i$, and we fix the hopping amplitude to $t=1$.
The second term in Eq.~(\ref{eq_ha}) introduces a random onsite potential (i.e., disorder) with strength $W$, and $r_i$ are identically and independently distributed random numbers in the interval $[-1,1]$. 
We consider both open boundary conditions (OBC) and periodic boundary conditions (PBC).

The single-particle spectrum of the Anderson model is bounded within $[-W/2-2,W/2+2]$. For finite systems and strong disorders, the single-particle energy levels, $\epsilon_\alpha$, are uncorrelated. For finite systems and weak disorders, they exhibit pairwise degeneracy for PBC, see Fig.~\ref{fig_1}(a), and pairwise equidistance for OBC, see Fig.~\ref{fig_speca}(b). Therefore, they manifest the same behavior as the free fermion point, where $\epsilon_\alpha = -2\cos(2\pi k /L)$ for PBC and $\epsilon_\alpha = -2\cos(\pi k /(L+1))$ for OBC, where $k$ is an integer~\cite{PhysRevE.86.011142,PhysRevE.100.022142}. 
Generally, the single-particle energy eigenstates, $|\alpha\rangle$, are exponentially localized with the localization length away from the edges and middle of the spectrum,
\begin{equation}
\label{eq:zeta1}
    \zeta_\text{A}\approx \frac{96(1-\epsilon_\alpha^2/4)}{W^2},
\end{equation}
and in the middle of the spectrum,
\begin{equation}
\label{eq:zeta2}
    \zeta_\text{A}\approx\frac{\Gamma(1/4)^2}{\Gamma(3/4)^2} \frac{12}{W^2},
\end{equation}
where $\Gamma$ stands for the gamma function~\cite{Thouless1983,Thouless1979,Thouless1974,Kappus1981,Tessieri_2012}.
Consequently, the Anderson model is localized for an arbitrary disorder strength $W>0$ in the thermodynamic limit $L\rightarrow\infty$.

The second model is the Wannier-Stark model, which describes a 1D lattice with a tilted potential,
\begin{equation}
\label{eq_hws}
    \hat{H}_\text{W-S}=-t\sum_{i} \left(\hat{c}_i^\dagger \hat{c}_{i+1}+\text{h.c.}\right)+F\sum_{i} i \hat{c}_i^\dagger \hat{c}_{i}\,.
\end{equation}
The first term in Eq.~(\ref{eq_hws}) corresponds to the kinetic energy, and we fix the hopping amplitude to $t=1$. 
The second term in Eq.~(\ref{eq_hws}) introduces a tilted potential (i.e., a constant electric field) with strength $F$.
In this model we only consider open boundary conditions (OBC).

The single-particle spectrum of the Wannier-Stark model away from the spectral edges forms an ordered ladder with $\epsilon_\alpha=F\alpha$~\cite{PhysRevB.8.5579}.
The single-particle energy eigenstates can be expressed as $|\alpha\rangle=\sum_{i}\mathcal{J}_{i-\alpha}(2/F) |i\rangle$, where ${J}_{i-\alpha}$ stands for the Bessel function of the first kind.
Moreover, $|\alpha\rangle$ is centered at site $\alpha$ with the localization length $\zeta_\text{WS} = 2/F$, so that the Wannier-Stark model is localized for an arbitrary field strength $F>0$ in the thermodynamic limit $L\rightarrow\infty$~\cite{van_Nieuwenburg_2019}.

In this manuscript, we focus on the properties of typical energy eigenstates, which lie away from the tails of the density of states. It was previously shown that the fraction of atypical energy eigenstates vanishes in the large system size limit~\cite{Vidmar_2017}. Accordingly, all numerical results are averaged over a finite number of energy eigenstates from the middle of the spectrum. We emphasize that, for a vanishing rescaled on-site potential, the density of states develops peaks near the spectral edges. However, these features disappear and the density of states becomes featureless as the scaled on-site potential increases toward the eigenstate transition in both the 1D Anderson and Wannier-Stark models. This is demonstrated in App.~\ref{Appendix_DOS}.

\section{Eigenstate transitions in nonstandard thermodynamic limit} \label{sec:nonstandard}

While the discussion in the previous section suggests that the single-particle energy eigenstates are localized in position space for any nonzero $W$ and $F$, we here introduce the scaled parameters for on-site potentials, $\tilde{W}$ and $\tilde{F}$, such that the transition to position-space localization occurs at some nonzero $\tilde{W}$ and $\tilde{F}$.
This approach, termed nonstandard thermodynamic limit, has been recently pursued in several works on disordered single-particle~\cite{PhysRevE.100.022142,PhysRevE.86.011142} and many-body~\cite{PhysRevB.97.214205,Gopalakrishnan_2019,PhysRevB.109.214203,PhysRevB.108.174303,PhysRevB.105.214308} systems.

We emphasize that the delocalized (localized) regime refers to the regime where the localization length in the middle of the spectrum is larger (smaller) than the system size. Hence, scaling with $W$ describes what happens when the system size $L$ increases while the disorder strength $W$ remains fixed. In this scenario, even if the Anderson or Wannier-Stark system starts in the delocalized regime, its localization length eventually becomes smaller than the number of lattice sites. Conversely, scaling with $\tilde{W}$ captures the behavior when the system size $L$ increases while the disorder strength $W$ decreases in such a way that a sample, which is initially in the delocalized regime, remains in that regime.

Our goal is to introduce a scaled parameter such that a fixed value of the latter implies a fixed ratio of the localization length and the system size, $\zeta/L = \rm const$.
In the Anderson model, this is achieved by
\begin{equation} \label{def_tildeW}
    \frac{\zeta_\text{A}}{L} \propto \frac{1}{W^2 L} = {\rm const.}\;\;\; \to \;\;\; \tilde{W} = W \sqrt{L} \;,
\end{equation}
while in the Wannier-Stark model, this is achieved by
\begin{equation} \label{def_tildeF}
    \frac{\zeta_{\rm WS}}{L} \propto \frac{1}{F L} = {\rm const.}\;\;\; \to \;\;\; \tilde{F} = F L\;.
\end{equation}
For the Wannier-Stark model, an alternative way of introducing the scaled parameter $\tilde{F}$ is to renormalize the tilted potential operator, cf.~the second term on the right-hand side~of Eq.~(\ref{eq_hws}).
Namely, the operator of the tilted potential is super extensive, so that it has a diverging Hilbert-Schmidt norm when compared to the hopping operator. For the definition of the Hilbert-Schmidt norm, see Ref.~\cite{lydzba_swietek_24}.
The extensive counterpart of the tilted potential operator is $(1/L)\sum_i\, i \hat n_i$, implying that the effective model parameter is $\tilde{F}$ from Eq.~(\ref{def_tildeF}).

In the remaining part of this section, we illustrate the difference between the standard and nonstandard thermodynamic limits in both models through the lens of quantum dynamics and the structure of energy eigenstates.

\subsection{Single-particle quantum dynamics}

A convenient approach to distinguish distinct time and energy regimes in the models under investigation is to study the quantum dynamics.
We consider the quantum quench from an initial single-particle state $|\psi_0\rangle= |i\rangle=\hat c_i^\dagger |\emptyset\rangle$, in which the particle is localized on site $i$ in the middle of the lattice.
We define the width $\Sigma(t)$ of the time-evolving wave packet as
\begin{equation}
    \Sigma(t)^2 = \sum_{j=1}^{L} (j-i)^2\langle \psi(t)| \hat n_j |\psi(t)\rangle \;,
\end{equation}
where $|\psi(t)\rangle = e^{-i\hat H t}|\psi_0\rangle$, setting $\hbar\equiv 1$.
We model the dynamics with the ansatz $\Sigma(t) = a_0 \, t^{1/z}$, where $a_0$ is a constant and the dynamical exponent $z$ is expected to be $z=1$ for ballistic dynamics and $1/z=0$ for localization in the site occupation basis.

\begin{figure}[!t]
\includegraphics[width=\columnwidth]{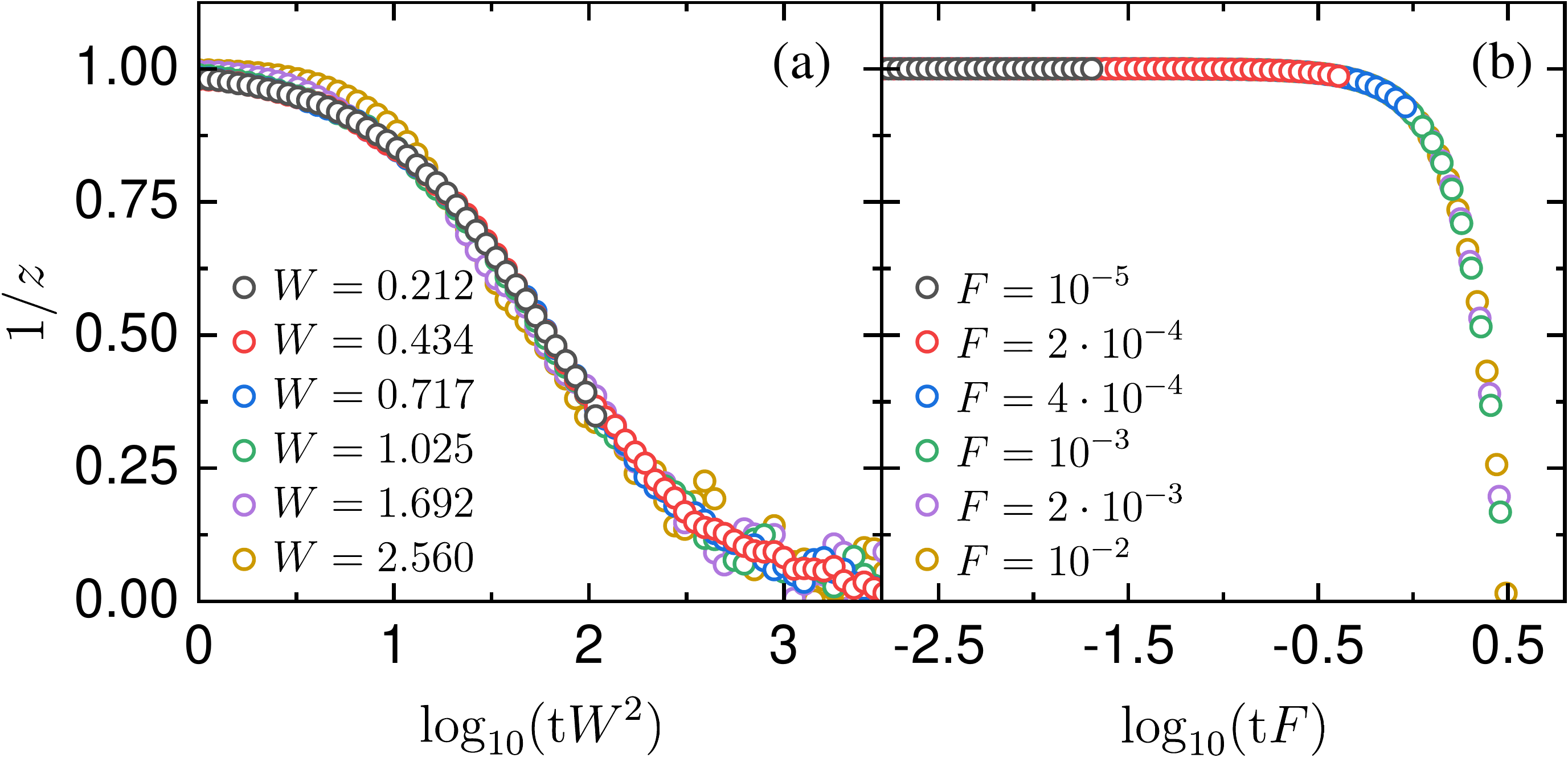}
\caption{
Time evolution of the inverse dynamical exponent $z(t)^{-1}$ versus the logarithm of the scaled time $\tilde t$.
(a) $\tilde t = tW^2$ in the Anderson model and (b) $\tilde t = tF$ in the Wannier-Stark model.
The values of potential strengths $W$ and $F$ are indicated in the panels, and the system size is $L = 10000$. Numerical results in (a) were averaged over $100$ disorder realizations.}
\label{fig_t}
\end{figure}

In the actual time evolution, the width of the wave packet does not necessarily behave as $\Sigma(t) \propto t^{1/z}$ for all times, which can be interpreted as the dynamical exponent $z$ evolving in time.
We hence define the temporal inverse of the dynamical exponent as
\begin{equation} \label{def_z_inverse}
    z(t)^{-1} = \frac{d \log \Sigma(t)}{d \log t}\;.
\end{equation}
Results for $z(t)^{-1}$ are shown against the scaled time $tW^2$ in the Anderson model, see Fig.~\ref{fig_t}(a), and against the scaled time $tF$ in the Wannier-Stark model, see Fig.~\ref{fig_t}(b).
In both cases, we observe a collapse of the results at different values of $W$ and $F$.
At short times, $z =1$ to high precision, suggesting ballistic transport, while at long times, $z^{-1}\to 0$, suggesting localization in position space.
In the intermediate regime, $1/z$ smoothly evolves from 1 to 0, indicating an absence of a transient diffusive transport~\cite{Bertini_heidrichmeisner_21}.

Since a quantum-chaotic regime is absent in the models under investigation, the characteristic relaxation time, denoted as the Thouless time $t_{\rm Th}$, cannot be defined in the conventional way (e.g., via the spectral form factor~\cite{Suntajs_2021}). Instead, we argue that a convenient way to extract the Thouless time is via the wave packet spreading studied in Fig.~\ref{fig_t}.
We define the Thouless time as the characteristic time of the crossover from the ballistic dynamics ($z=1$) to localization in position space ($1/z=0$).
This occurs at $t_{\rm Th}W^2 = {\rm const.}$ in the Anderson model, see Fig.~\ref{fig_t}(a), and at $t_{\rm Th}F = {\rm const.}$ in the Wannier-Stark model, see Fig.~\ref{fig_t}(b).
Hence, $t_{\rm Th}\propto 1/W^2$ in the Anderson model and $t_{\rm Th} \propto 1/F$ in the Wannier-Stark model.
These scalings suggest that in the standard thermodynamic limit, the wave packets would localize in a finite time for any nonzero $W$ and $F$.

The nonstandard thermodynamic limit can be introduced by setting up the criterion for the localization transition in the time domain~\cite{suntajs_bonca_20a, sierant_delande_20, Suntajs_2021}.
In this view, the transition occurs when the Thouless time $t_{\rm Th}$ equals the Heisenberg time $t_{\rm H}$, which is inversely proportional to the mean level spacing $\Delta$ of the single-particle spectrum, i.e., $t_{\rm H}\propto 1/\Delta \propto L$.
Hence, the characteristic disorder $W^*$ for the localization transition in the Anderson model scales as
\begin{equation}
    t_{\rm Th}(W=W^*) \propto t_H \propto L\;\;\; \to \;\;\; W^* \propto \frac{1}{\sqrt{L}}\;,
\end{equation}
and the characteristic on-site potential $F^*$ for the localization transition in the Wannier-Stark model scales as
\begin{equation}
    t_{\rm Th}(F=F^*) \propto t_H \propto L\;\;\; \to \;\;\; F^* \propto \frac{1}{L}\;.
\end{equation}
These scalings suggest that the transition to position-space localization occurs in the nonstandard thermodynamic limit as discussed above, characterized by the scaled parameters $\tilde{W}$ and $\tilde{F}$ introduced in Eqs.~(\ref{def_tildeW}) and~(\ref{def_tildeF}), respectively.

\subsection{Structure of single-particle eigenstates} \label{sec:structure_eigenstates}

Another perspective on the properties of models under investigation is obtained via the measures of localization or delocalization of Hamiltonian eigenfunctions on a given basis.
A standard measure of (de)localization of a single-particle energy eigenstate $|\alpha\rangle$ is the inverse participation ratio in the position space~\cite{Thouless1974},
\begin{equation}
\label{eq_ipr}
    \text{IPR}^{(i)} = \sum_{i=1}^{L}|\langle i | \alpha \rangle |^4\;,
\end{equation}
where $|i\rangle$ is the single-particle eigenstate of the site occupation operator $\hat{n}_{i}=\hat{c}_{i}^\dagger\hat{c}_{i}$. It determines the average number of sites with a nonvanishing amplitude $\langle i | \alpha \rangle$, so the average support of $|\alpha\rangle$. Since we are interested in the statistical properties of $|\alpha\rangle$ in the middle of the single-particle spectrum, we define
\begin{equation} \label{def_ipr_i}
    \text{ipr}^{(i)}=L\langle\langle \text{IPR}^{(i)} \rangle_\alpha\rangle_H\;,
\end{equation} 
with $\langle...\rangle_\alpha$ and $\langle ...\rangle_H$ standing for the averaging over single-particle energy eigenstates and disorder realizations, respectively. Note that $\langle ...\rangle_H$ is performed only for the~1D Anderson model. We note that other measures of (de)localization, not studied here, can also be considered. For example, the single-particle Rényi entropy, particularly well suited for studying the multifractal properties of crossovers in quadratic systems~\cite{PhysRevB.86.134201}, may represent a promising avenue for future research. 

We consider $|\alpha\rangle$ to be delocalized in the position space when $\text{ipr}^{(i)}$ is independent of the system size $L$. We categorize $|\alpha\rangle$ as completely delocalized (cD) when $\text{ipr}^{(i)} \approx 1$ and as partially delocalized (pD) when $\text{ipr}^{(i)} \gg 1$. Conversely, we consider $|\alpha\rangle$ to be localized in the position space when $ \text{ipr}^{(i)}$ is proportional to the system size $L$. Specifically, we define $|\alpha\rangle$ as completely localized (cL) when $\text{ipr}^{(i)} \approx L$ and as partially localized (pL) when $\text{ipr}^{(i)} \approx a_1 L$, with $a_1 \ll 1$.

\begin{figure}[t!]
\includegraphics[width=\columnwidth]{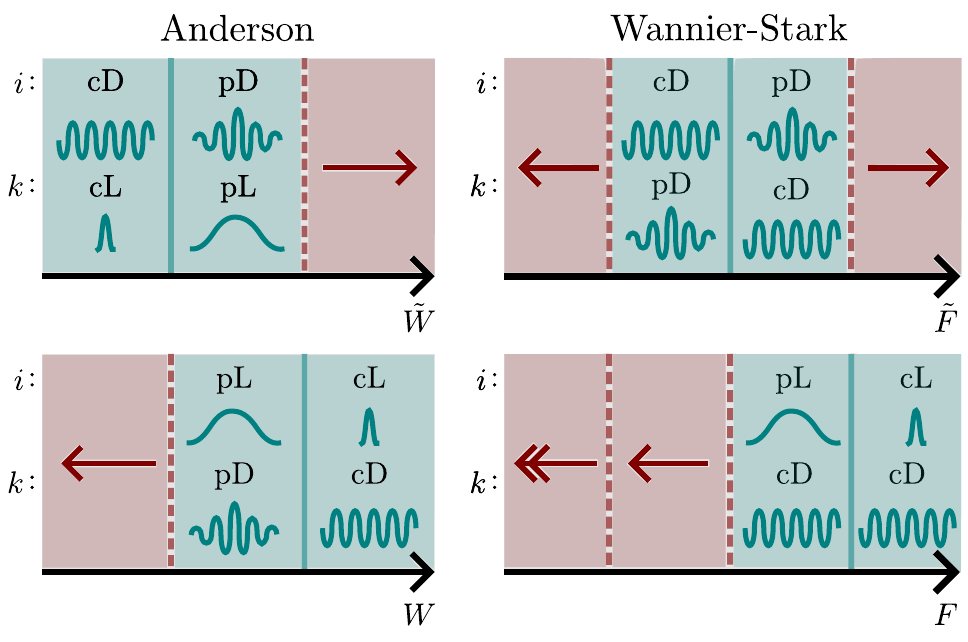}
\caption{The sketch of eigenstate transitions in the Anderson and Wannier-Stark models under both the nonstandard ($\tilde{W}$ and $\tilde{F}$) and standard ($W$ and $F$) thermodynamic limits. Red regimes vanish in these limits with their dependencies on the system size indicated by red arrows. Simultaneously, green regimes do not vanish in these limits and are labeled as pD (partial delocalization), cD (complete delocalization), pL (partial localization), and cL (complete localization). Wiggly and nonwiggly curves symbolize delocalization and localization, respectively. They do not depict the position or quasimomentum wave functions. It is apparent that the eigenstate transitions in the Anderson and Wannier-Stark models have similar properties in the position space ($i$) but different in the quasimomentum space ($k$). This is further confirmed by the studies of entanglement entropies in Sec.~\ref{sec_ee}.}
\label{fig_sketch}
\end{figure}

It is also beneficial to analyze the inverse participation ratio in the quasimomentum space,
\begin{equation}
     \text{IPR}^{(k)} = \sum_{k=1}^{L}|\langle k | \alpha \rangle |^4\;,
\end{equation} 
where $|k\rangle$ is the single-particle eigenstate of the quasimomentum occupation operator $\hat{m}_{k}=\frac{1}{L}\sum_{i,j=1}^{L} e^{-\text{i}\frac{2\pi}{L} k(i-j)}\hat{c}_{i}^\dagger \hat{c}_{j}$. In analogy to Eq.~(\ref{def_ipr_i}), we define
\begin{equation} \label{def_ipr_k}
    \text{ipr}^{(k)}=\frac{3}{2}L\langle\langle \text{IPR}^{(k)} \rangle_\alpha\rangle_H\;.
\end{equation} 
Note the different coefficients in the definitions of $\text{ipr}^{(i)}$ and $\text{ipr}^{(k)}$. This difference ensures that $\text{ipr}^{(i)}$ exhibits the same value in the limit of small scaled potentials ($\tilde{W}, \tilde{F} \to 0$), as $\text{ipr}^{(k)}$ in the limit of large scaled potentials ($\tilde{W}, \tilde{F} \to \infty$).

\begin{figure}[t!]
\includegraphics[width=\columnwidth]{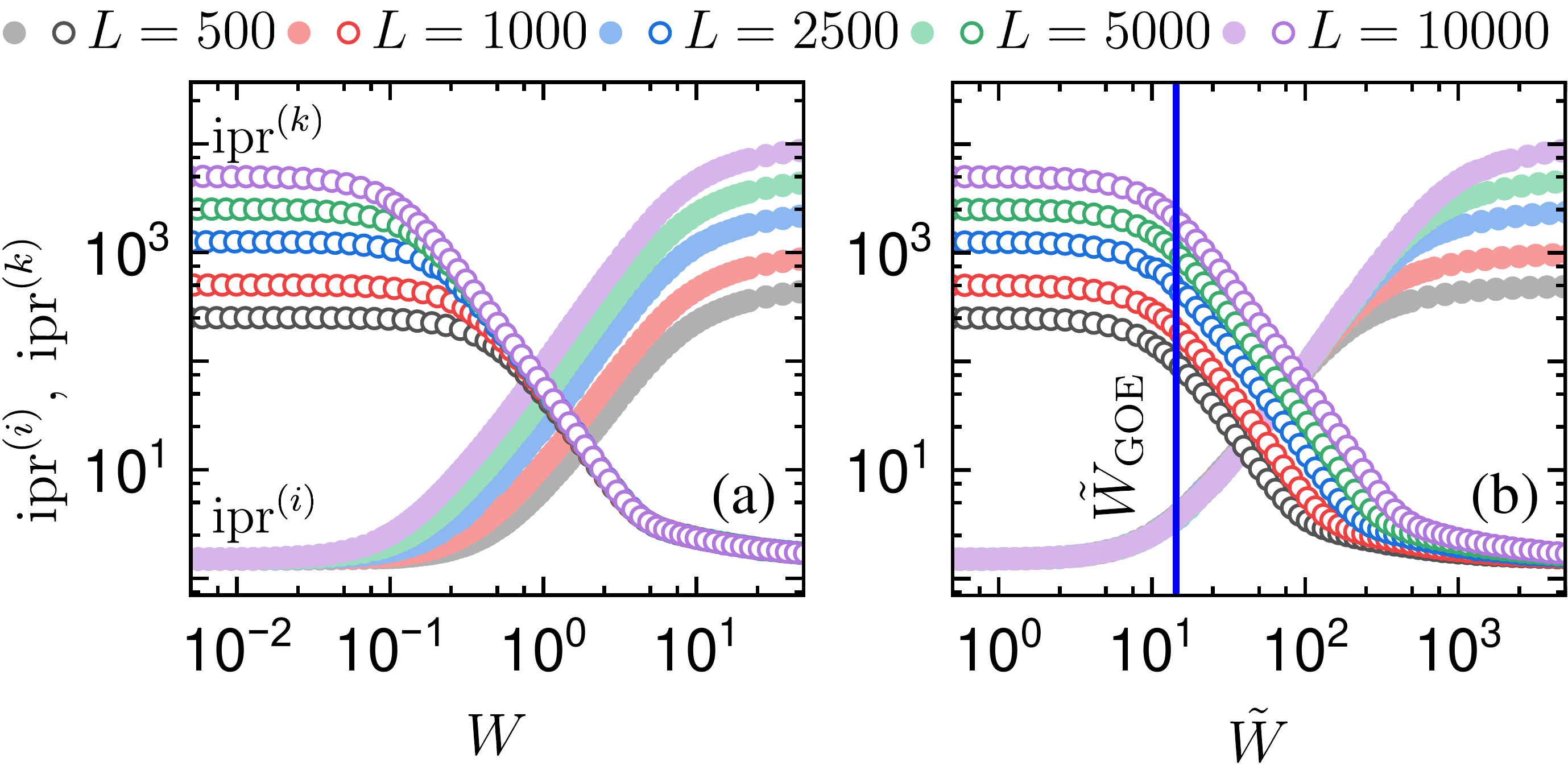}
\caption{The inverse participation ratios $\text{ipr}^{(i)}$ and $\text{ipr}^{(k)}$, see Eqs.~(\ref{def_ipr_i}) and~(\ref{def_ipr_k}), plotted as a function of (a)~$W$ and (b)~$\tilde{W}$ in the Anderson model, respectively. We consider PBC and $L=500, 1000, 2500, 5000$ and $10000$. Filled symbols correspond to $\text{ipr}^{(i)}$, while empty symbols correspond to $\text{ipr}^{(k)}$. The blue vertical line in (b) marks the scaled disorder strength $\tilde{W}_{\rm GOE}$, for which the spectral statistics agree with the GOE prediction, see App.~\ref{Appendix_spectral} and Fig.~\ref{fig_1}. Numerical results were averaged over $100$ single-particle energy eigenstates in the middle of the spectrum and $20$ disorder realizations. Legend provided above the figure is valid for all panels.}
\label{fig_ipr1}
\end{figure}

\begin{figure}[t!]
\includegraphics[width=\columnwidth]{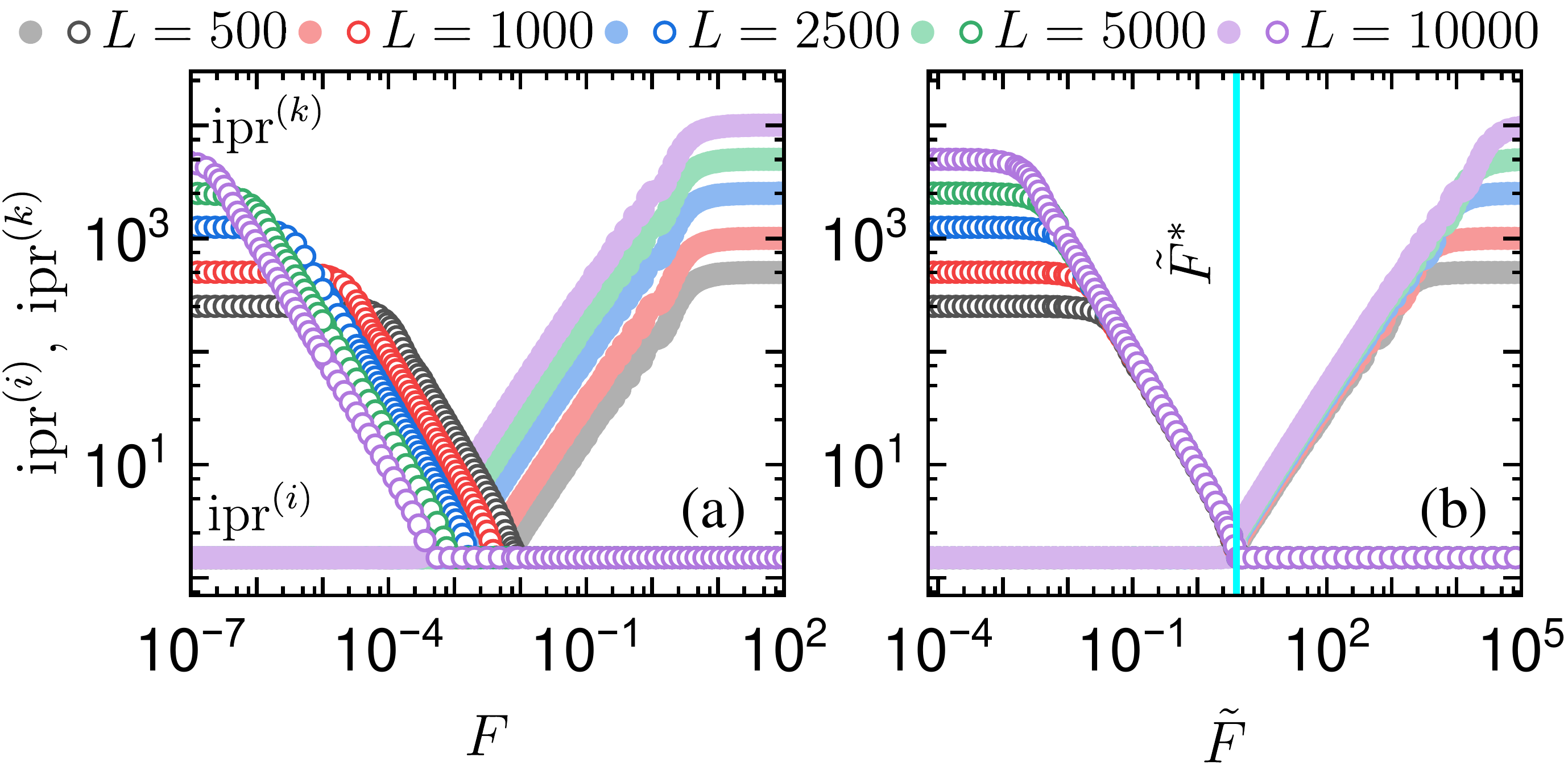}
\caption{The inverse participation ratios $\text{ipr}^{(i)}$ and $\text{ipr}^{(k)}$, see Eqs.~(\ref{def_ipr_i}) and~(\ref{def_ipr_k}), plotted as a function of (a)~$F$ and (b)~$\tilde{F}$ in the Wannier-Stark model, respectively. We consider $L=500, 1000, 2500, 5000$ and $10000$. Filled symbols correspond to $\text{ipr}^{(i)}$, while empty symbols correspond to $\text{ipr}^{(k)}$. The cyan vertical line in (b) marks the sudden jump in the inverse participation ratio, see App.~\ref{Appendix_ipr}. Numerical results were averaged over $100$ single-particle energy eigenstates in the middle of the spectrum. Legend provided above the figure is valid for all panels.}
\label{fig_ipr2}
\end{figure}

We first present a sketch of the eigenstate transitions in the Anderson and Wannier-Stark models under both the nonstandard and standard thermodynamic limits, see Fig.~\ref{fig_sketch}, while detailed calculations of $\text{ipr}^{(i)}$ and $\text{ipr}^{(k)}$ are shown in Figs.~\ref{fig_ipr1} and~\ref{fig_ipr2}. 

Interestingly, the eigenstate properties of the Anderson and Wannier-Stark models are similar in the position space, see the upper rows (denoted by $i$) in the sketches in Fig.~\ref{fig_sketch}. In the nonstandard thermodynamic limit, the transition occurs between the regimes of cD and pD. In the standard thermodynamic limit, it occurs between pL and cL. In contrast, the properties of these models differ in the quasimomentum space, see the lower rows (denoted by $k$) in the sketches in Fig.~\ref{fig_sketch}. In the Anderson model, the single-particle energy eigenstates transition from cL to pL with the increasing $\tilde{W}$, while from pD to cD with the increasing $W$. In the Wannier-Stark model, the single-particle energy eigenstates transition from pD to cD with the increasing $\tilde{F}$, while remain cD across the entire range of $F$. An analogous similarity (dissimilarity) in position space (quasimomentum space) is also observed in the entanglement entropies of typical many-body energy eigenstates, see Sec.~\ref{sec_ee}.

\begin{figure}[t!]
\includegraphics[width=\columnwidth]{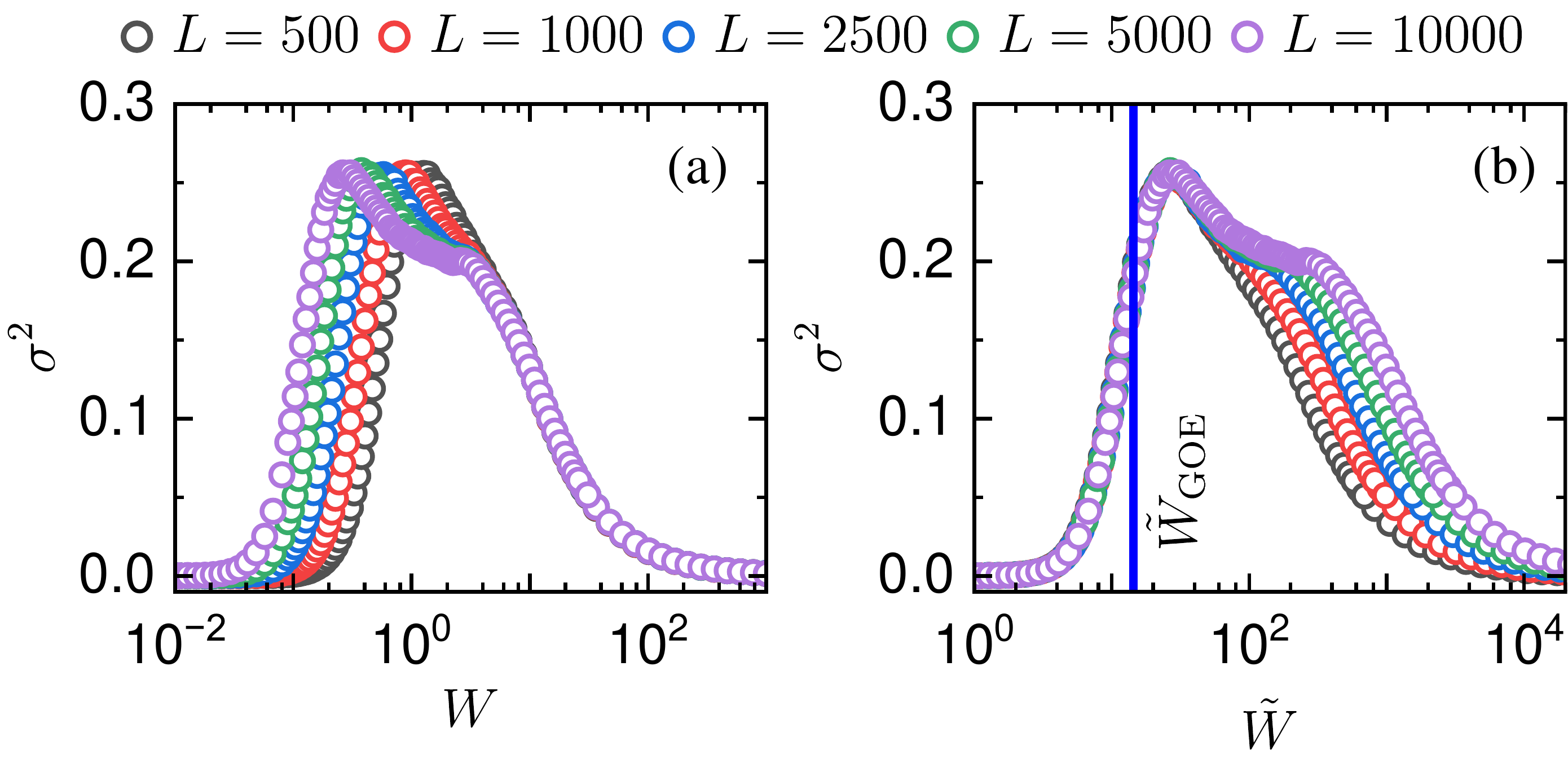}
\caption{The relative variance $\sigma^2$, see Eq.~(\ref{def_variance_sigma}), versus (a) $W$ and (b) $\tilde{W}$ in the Anderson model with PBC. We consider $L=500, 1000, 2500, 5000$ and $10000$. The blue vertical line in (b) marks the scaled disorder strength $\tilde{W}_{\rm GOE}$ for which the spectral statistics agree with the GOE prediction, see App.~\ref{Appendix_spectral} and Fig.~\ref{fig_1}. Numerical results were averaged over $100$ single-particle energy eigenstates in the middle of the spectrum and $10000$ disorder realizations. Legend is provided above the figure.}
\label{fig_ipr3}
\end{figure}

In Fig.~\ref{fig_ipr1}, we present results of numerical calculations of $\text{ipr}^{(i)}$ and $\text{ipr}^{(k)}$ in the Anderson model. We only include the results for PBC, as the ones for OBC are qualitatively and quantitatively similar. Additionally, the corresponding results for the Wannier-Stark model are illustrated in Fig.~\ref{fig_ipr2}. The behaviour of $\text{ipr}^{(i)}$ and $\text{ipr}^{(k)}$ in the opposing limits of large and small scaled parameter strengths is the same in both models. When $\tilde{W}$ and $\tilde{F}$ are large, $\text{ipr}^{(i)}=L$ and $\text{ipr}^{(k)}=\frac{3}{2}$. When $\tilde{W}$ and $\tilde{F}$ are small, $\text{ipr}^{(i)}=\frac{3}{2}$ and $\text{ipr}^{(k)}=\frac{L}{2}$. The latter agrees with the prediction for translationally invariant free fermions, i.e., $\text{ipr}^{(i)} = \frac{3}{2}$~\cite{PhysRevE.100.022142}. This is specific to the one-dimensional version of the Anderson model. In higher dimensions~\cite{Schubert2005,PhysRevB.107.064205,Suntajs_2021}, the inverse participation ratio actually agrees with the GOE prediction, i.e., $\text{ipr}^{(i)} = 3$~\cite{ZELEVINSKY_1996,dalessio_kafri_16}. This is most likely related to the fact that in higher dimensions, the singe-particle energy eigenstates can be simultaneously delocalized in position and quasimomentum spaces, allowing for the model to support the single-particle quantum chaos~\cite{Lydzba_2021,Lydzba_2021b}.
 
The differences between the models are revealed in the intermediate regime of $\tilde{W}$ and $\tilde{F}$. The eigenstate transition in the Anderson model is signaled by a slow increase of $\text{ipr}^{(i)}$ and slow decrease of $\text{ipr}^{(k)}$, see Fig.~\ref{fig_ipr1}. Its position is roughly captured by $\tilde{W}_\text{GOE}=14.36$, which we determined from the spectral statistics, see Fig.~\ref{fig_1} and App.~\ref{Appendix_spectral} for details.
We shall use this value of $\tilde{W}_\text{GOE}$ as the reference point in further calculations.

The nature of the eigenstate transition can also be characterized using the relative variance,
\begin{equation} \label{def_variance_sigma}
    \sigma^2 = \left<\frac{\langle x^2 \rangle_\alpha -\langle x \rangle_\alpha^2 }{\langle x \rangle_\alpha^2}\right>_H \;,
\end{equation}
with $x=\text{IPR}^{(i)}$. When all single-particle energy eigenstates are cD or cL, this variance is expected to be negligibly small. Since the localization length depends on the single-particle energy in the Anderson model, see Eqs.~(\ref{eq:zeta1}) and (\ref{eq:zeta2}), it is natural to expect that the variance increases when $\zeta_\text{A}$ of the first single-particle energy eigenstates becomes smaller than the system size, it exhibits a maximum marking of the eigenstate transition, and then decreases when $\zeta_\text{A}$ of the last single-particle energy eigenstates becomes smaller than the system size. We verify this surmise in Fig.~\ref{fig_ipr3}. Indeed, there is a feature in $\sigma^2$ with a maximum close to $\tilde{W}_\text{GOE}=14.36$. Its left shoulder is scale invariant in $\tilde{W}$, while its right shoulder is scale invariant in $W$. This is not unexpected, since the regime of cL (cD) vanishes in the nonstandard (standard) thermodynamic limit.

In the Wannier-Stark model, the eigenstate transition is marked by an abrupt change in $\text{ipr}^{(i)}$ and $\text{ipr}^{(k)}$, indicating the shift from cD in position space to cD in quasimomentum space, see Fig.~\ref{fig_ipr2}. This allows us to numerically establish the critical scaled field strength, i.e, $\tilde{F}^*\approx 4.00$. More details can be found in App.~\ref{Appendix_ipr}.

\section{Single-particle spectral statistics} \label{sec_ss}

The statistics of the nearest level spacings $\delta_\alpha=\epsilon_{\alpha}-\epsilon_{\alpha-1}$ is commonly used as a measure of quantum chaos. In order to avoid the spectral unfolding~\cite{PhysRevE.84.016203,Abuelenin_2018}, it is convenient to study the ratio 
\begin{equation}
\label{eq_ratio}
    \tilde{r} = \frac{\text{min}(\delta_{\alpha+1},\delta_\alpha)}{\text{max}(\delta_{\alpha+1},\delta_\alpha)},
\end{equation} 
which is restricted to the interval $[0,1]$. When the model is chaotic, the energy eigenstates are correlated and avoid crossings, so that the distribution of their spacings agrees with the GOE prediction, while $\langle\langle\tilde{r}\rangle_\alpha\rangle_H$ is close to $\tilde{r}_\text{GOE}\approx 0.536$~\cite{PhysRevB.75.155111,PhysRevLett.110.084101}. In contrast, when the model is not chaotic, the energy eigenstates are typically uncorrelated, leading to the Poisson distribution of spacings, and $\langle\langle\tilde{r}\rangle_\alpha\rangle_H$ matching $\tilde{r}_\text{P}\approx 0.386$~\cite{PhysRevB.75.155111,PhysRevLett.110.084101}. We note that in the latter case, the spectral statistics may also exhibit a picket fence-like structure, characterized by one or more pronounced maxima~\cite{Fogarty_2021,PhysRevA.106.013301,PhysRevA.43.4237}.

The spectral statistics of the Anderson model with OBC has been already studied in Refs.~\cite{PhysRevE.86.011142,PhysRevE.100.022142}. For small $\tilde{W}$, the mean ratio aligns with the prediction for pairwise equidistant energy levels, $\langle\langle\tilde{r}\rangle_\alpha\rangle_H\approx 1$. For large $\tilde{W}$, the mean ratio matches the expectation for uncorrelated energy levels, $\langle\langle\tilde{r}\rangle_\alpha\rangle_H\approx \tilde{r}_\text{P}$. Moreover, $\langle\langle\tilde{r}\rangle_\alpha\rangle_H$ is a smooth function of $\tilde{W}$, so that it crosses $\tilde{r}_\text{GOE}$ at $\tilde{W}_\text{GOE}\approx14.36$, see App.~\ref{Appendix_spectral} for details. 

In Refs.~\cite{PhysRevE.86.011142,PhysRevE.100.022142}, it has been argued that $\tilde{r}_\text{GOE}$, exhibited near the eigenstate transition, is not a signature of single-particle quantum chaos. It is a mere consequence of single-particle energy levels experiencing all degrees of repulsion when going from the ladder structure to the uncorrelated spectrum. This interpretation is supported by the observation that $\langle\langle\tilde{r}\rangle_\alpha\rangle_H\approx \tilde{r}_\text{GOE}$ only at $\tilde{W}_\text{GOE}$. In higher dimensions, where the Anderson model supports single-particle quantum chaos, this agreement is observed over a wide range of disorder strengths~\cite{PhysRevB.95.094204, Suntajs_2021, PhysRevB.107.064205}.

Another perspective may be suggested by the spectral statistics of the Anderson model with PBC, in which $\langle\langle\tilde{r}\rangle_\alpha\rangle_H$ exhibits a peak at $\tilde{W} \approx \tilde{W}_\text{GOE}$, as shown by Fig.~\ref{fig_1}(a).
Interestingly, the mean ratio at $\tilde{W} \approx \tilde{W}_\text{GOE}$ is $\langle\langle\tilde{r}\rangle_\alpha\rangle_H \approx \tilde{r}_\text{GOE}$, thereby coinciding with the mean ratio at $\tilde{W} = \tilde{W}_\text{GOE}$ for OBC, as studied in App.~\ref{Appendix_spectral}. 
A similar agreement with the GOE universality is observed for the entanglement entropy of typical many-body energy eigenstates with the bipartition in the quasimomentum space, as will be discussed in Sec.~\ref{sec_ee}.

Before concluding this section, we mention that in the Wannier-Stark model, $\langle\langle\tilde{r}\rangle_\alpha\rangle_H\approx 1$ across the whole range of $\tilde{F}$. Although there is a shallow minimum near the critical field strength $\tilde{F}^*$, it vanishes in the nonstandard thermodynamic limit, see App.~\ref{Appendix_spectral}.

\section{Entanglement entropy in many-body eigenstates}
\label{sec_ee}

\begin{figure*}[t!]
\includegraphics[width=\textwidth]{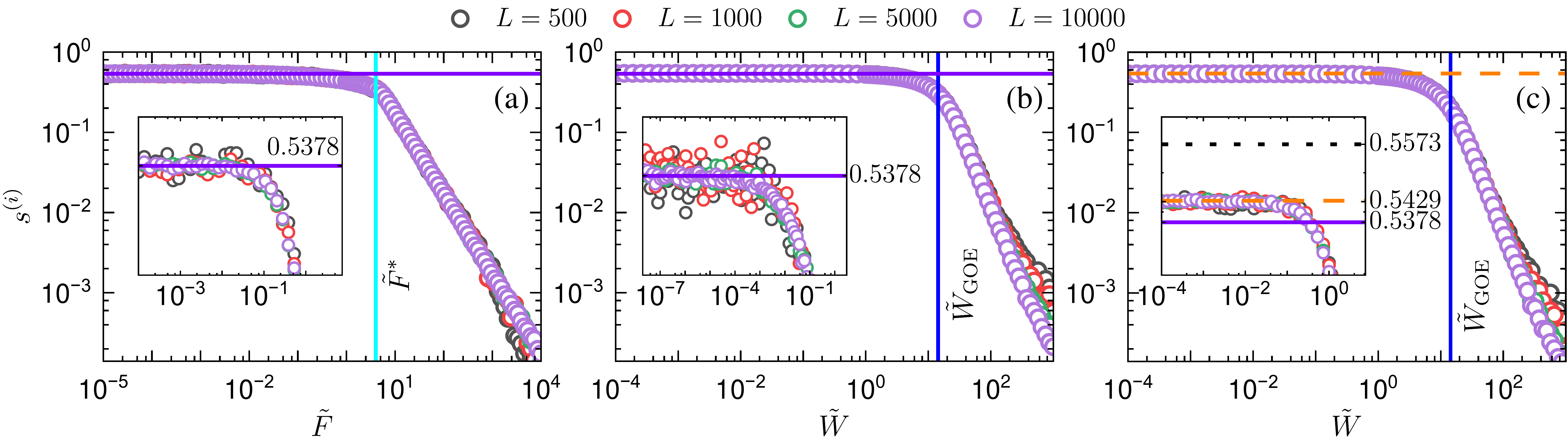}
\caption{The volume-law coefficient $s^{(i)}$ plotted versus the scaled potential strength. The scaled potential strength is (a) $\tilde{F}$ in the Wannier-Stark model, and (b),(c) $\tilde{W}$ in the Anderson model with OBC and PBC, respectively. Insets display close-ups near $\tilde{F},\tilde{W}\rightarrow 0$. The cyan vertical line in (a) marks $\tilde{F}^*$, while the blue vertical line in (b),(c) marks $\tilde{W}_\text{GOE}$. The violet solid horizontal line in (a)-(c) marks $s^{(i)}_\text{free}$, the black dotted horizontal line in (c) marks $s^{(i)}_\text{chaotic}$, while the orange dashed horizontal line marks the average value of $s^{(i)}$ in the interval $\tilde{W}\in[10^{-8},10^{-5}]$. Numerical results were averaged over $100$ many-body eigenstates in the middle of the spectrum. They were additionally averaged over $500$ disorder realizations in (b),(c). The legend provided above the figure is valid for all panels.}
\label{fig_e1}
\end{figure*}

\subsection{Bipartitions in position space} \label{sec:bipartitions_real}

Since our models are quadratic, the many-body energy eigenstates can be constructed as $|m\rangle=\prod_{\alpha\in\mathcal{N}} \hat{a}_\alpha^\dagger |\emptyset\rangle$, where $\hat{a}_\alpha^\dagger$ creates a spinless fermion in $|\alpha\rangle$, while $\mathcal{N}$ stands for a set of all occupied single-particle energy eigenstates. The corresponding many-body energies are $E_m=\sum_{\alpha\in\mathcal{N}} \epsilon_\alpha$. We consider the sector at half filling, i.e., $n=\frac{N}{L}=\frac{1}{2}$. In order to calculate the von Neumann entanglement entropy of an eigenstate $|m\rangle$, we bipartition it into two connected subsystems, i.e., the subsystem A with $L_A$ contiguous lattice sites and the environment B with $L-L_A$ lattice sites. We define the subsystem fraction as $f=\frac{L_A}{L}$, and we fix it to $f=\frac{1}{2}$. One could calculate the reduced density matrix of A by tracing out all of the degrees of freedom of B, i.e., $\hat{\rho}_\text{A}=\text{Tr}_\text{B}(|m\rangle\langle m|)$. The von Neumann entanglement entropy would be then $S^{(i)}=-\text{Tr}(\hat{\rho}_\text{A}\ln\hat{\rho}_\text{A})$, where the upper index $(i)$ denotes the bipartition in position space. In quadratic models, $S^{(i)}$ can also be evaluated from the eigenvalues $\lambda_{j}$ of the one-body correlation matrix~\cite{PhysRevLett.90.227902,PhysRevB.64.064412,Peschel_2003,Peschel_2009},
\begin{equation}
\label{eq_J}
    J^{(i)}_{lj}=2\langle m| \hat{c}_{l}^\dagger \hat{c}_{j} |m\rangle-\delta_{lj},
\end{equation}
as
\begin{equation}
    S^{(i)}=-\sum_{j=1}^{L_A} \left(\frac{1+\lambda_j}{2}\ln\left[\frac{1+\lambda_j}{2}\right]+\frac{1-\lambda_j}{2}\ln\left[\frac{1-\lambda_j}{2}\right]\right).
\end{equation}
In the above equations, $l,j\le L_A$ and $\lambda\in[-1,1]$.

It is well established that the average entanglement entropy $\langle\langle S^{(i)}\rangle_m\rangle_H$ of quantum-chaotic interacting Hamiltonians exhibits the volume law scaling of the leading order term, i.e.,
\begin{equation}
    \langle\langle S^{(i)}\rangle_m\rangle_H = L_A \ln(2),
\end{equation}
which is the same as for the thermodynamic entropy at infinite temperature~\cite{Bhakuni_2020,PhysRevB.101.060401,PhysRevE.100.022131,PhysRevLett.119.220603,PhysRevE.87.042135,Deutsch_2010,PhysRevX.8.021026,Nakagawa:2017yiw}, and agrees with the entanglement entropy averaged over Haar-random pure states in the Hilbert space, known as the Page result~\cite{PhysRevLett.71.1291}. Simultaneously, the leading order term of $\langle\langle S^{(i)}\rangle_m\rangle_H$ in quadratic and integrable interacting Hamiltonians, provided that there is no localization in position space, still supports the volume law scaling,
\begin{equation}
    \langle\langle S^{(i)}\rangle_m\rangle_H = s^{(i)}(f) L_A \ln(2),
\end{equation}
but the volume law coefficient is not maximal for a nonvanishing subsystem fraction, i.e., $s^{(i)}(f)<1$ for $f>0$~\cite{Beugeling_2015,PhysRevE.109.024117,PhysRevLett.121.220602,PhysRevB.99.075123,PhysRevE.89.012125,PRXQuantum.3.030201}.

In general, the volume law coefficient $s^{(i)}(f)$ can take an arbitrary value from the interval $(0,1)$. Nevertheless, only two values are repeatedly reported in the literature. The first one,
\begin{equation}
    s_\text{chaotic}^{(i)}(f)=1-\frac{1+f^{-1}(1-f)\ln(1-f)}{\ln(2)}\;,
\end{equation}
has been established using the random matrix theory for a single-particle Hamiltonian~\cite{PhysRevLett.125.180604}. It agrees with the volume law coefficient of the entanglement entropy averaged over Haar-random Gaussian states in the Hilbert space~\cite{PhysRevB.103.L241118}. Moreover, it is universal for quantum-chaotic quadratic Hamiltonians, like the tight-binding billiards~\cite{Ul_akar_2022} or the 3D Anderson model~\cite{Lydzba_2021}. Its value at $f=1/2$ is $s_\text{chaotic}^{(i)}(\frac{1}{2})\approx 0.5573$. 
On the other hand, there is currently no closed-form analytical expression for the volume law coefficient of $ \langle\langle S^{(i)}\rangle_m\rangle_H$ for translationally invariant free fermions.
The numerical value at $f=1/2$ is $s_\text{free}^{(i)}(\frac{1}{2})\approx 0.5378$~\cite{Vidmar_2017}.
This value is very similar to the one obtained in the 1D Aubry–André model at small quasiperiodic potentials, for which the eigenstates are localized in quasimomentum basis~\cite{PhysRevLett.125.180604}.
Since we focus on $f=\frac{1}{2}$, we make the subsystem fraction dependence implicit, $s^{(i)}(\frac{1}{2})\rightarrow s^{(i)}$, from now on.

We plot $s^{(i)}$ as functions of the scaled parameters in the Wannier-Stark model, the Anderson model with OBC and PBC in Fig.~\ref{fig_e1}(a), \ref{fig_e1}(b) and \ref{fig_e1}(c), respectively. It is apparent that in all cases, $s^{(i)}$ is almost independent of the scaled parameter strength below the critical point, i.e., $\tilde{F}<\tilde{F}^*$ or $\tilde{W}<\tilde{W}_\text{GOE}$, while it polynomially decreases with the scaled parameter strength above the critical point. Interestingly, the volume law coefficient $s^{(i)}$ saturates to the value established for translationally invariant free fermions, $s^{(i)}_\text{free}$, in the limit of small $\tilde{F}$ and $\tilde{W}$ when the OBC are imposed, see the insets of Figs.~\ref{fig_e1}(a) and~\ref{fig_e1}(b). Simultaneously, $s^{(i)}$ saturates to $s^{(i)}\approx 0.5429$, which is larger than $s^{(i)}_\text{free}$ but smaller than $s^{(i)}_\text{chaotic}$, in the limit of small $\tilde{W}$ when the PBC are imposed, see the inset of Fig.~\ref{fig_e1}(c).

\begin{figure}[!b]
\includegraphics[width=\columnwidth]{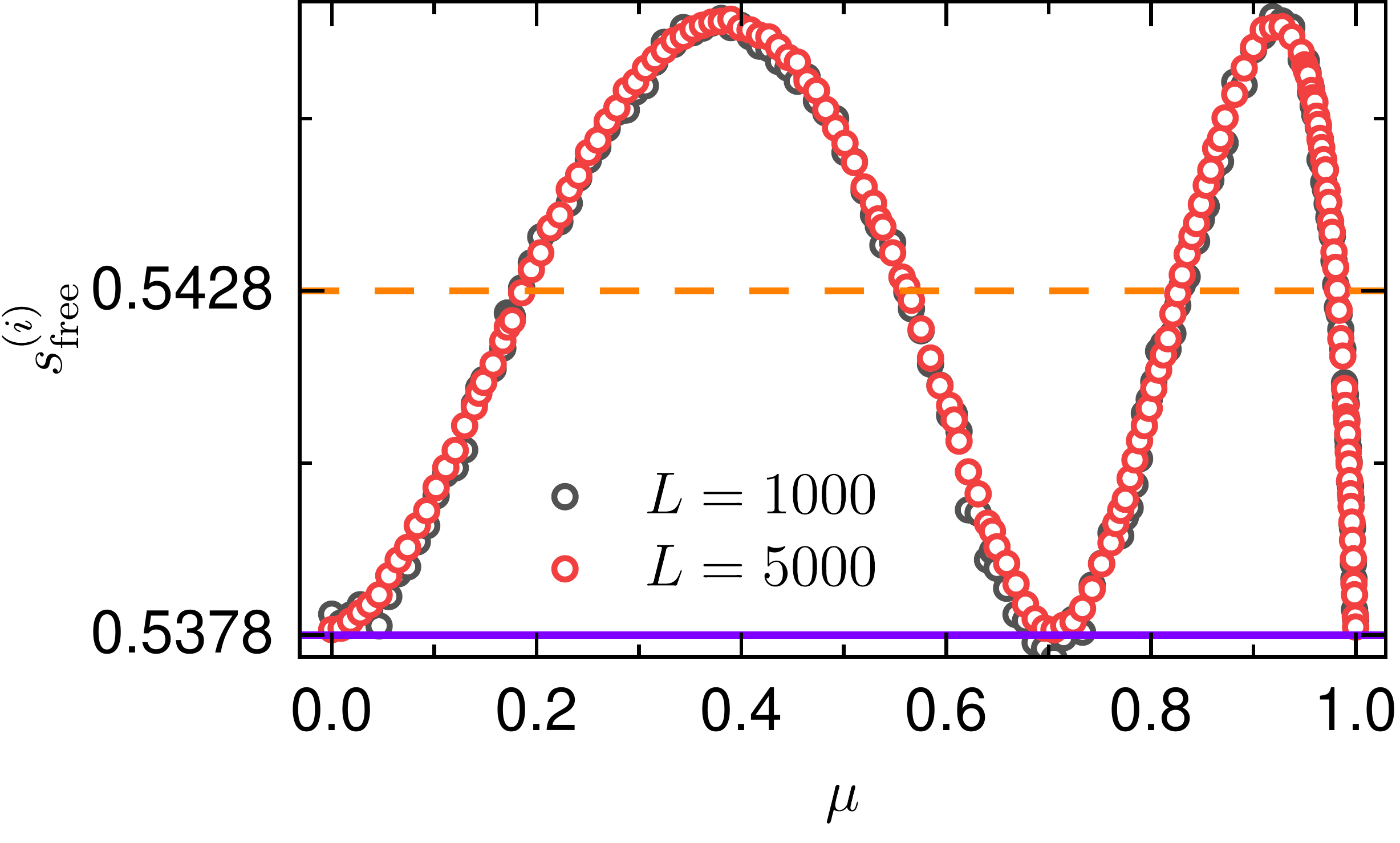}
\caption{The volume-law coefficient of translationally invariant free fermions, $s^{(i)}_\text{free}$, plotted versus the coupling of opposite quasimomenta, $\mu$. We consider two system sizes $L=100$ and $5000$. The violet solid line marks the previously published value, $s^{(i)}_\text{free}\approx0.5378$~\cite{Vidmar_2017}, while the orange dashed line marks the volume law coefficient averaged over couplings, $\langle s^{(i)}_\text{free}\rangle_\mu\approx 0.5428$. }
\label{fig_e2}
\end{figure}

\begin{figure*}[!t]
\includegraphics[width=\textwidth]{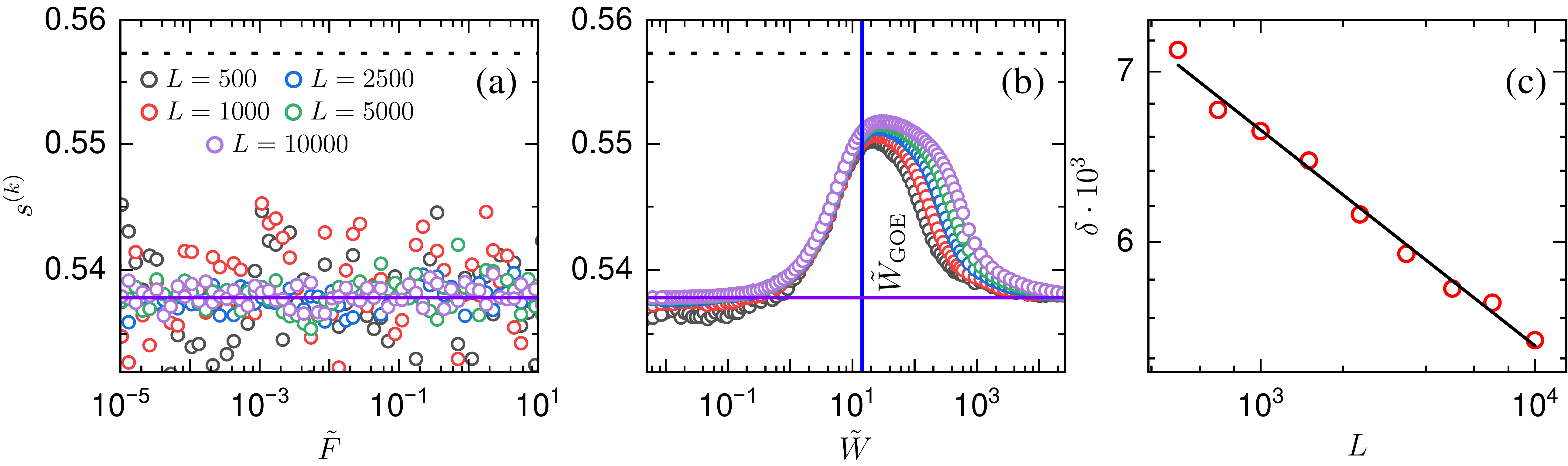}
\caption{The volume-law coefficient $s^{(k)}$ plotted against (a) $\tilde{F}$ in the Wannier-Stark model and (b) $\tilde{W}$ in the Anderson model with OBC. (c)~The finite-size scaling of the difference $\delta=s^{(i)}_\text{chaotic}-\text{max}[s^{(k)}]$ with the polynomial fit of $aL^b$ with $b\approx -0.085$. The blue solid vertical line in (b) marks $\tilde{W}_\text{GOE}$, the violet solid horizontal line in (a),(b) marks $s^{(i)}_\text{free}$, while the black dotted horizontal line in (b) marks $s^{(i)}_\text{chaotic}$. Numerical results were averaged over $100$ many-body eigenstates in the middle of the spectrum. They were additionally averaged over $500$ disorder realizations in (b),(c). Legend valid for (a),(b) panels is provided in the first panel of the figure.}
\label{fig_e3}
\end{figure*}

The absence of agreement between $s^{(i)}$ and $s^{(i)}_\text{free}$ below the localization-to-localization transition in the Anderson model with PBC can be explained as follows. The translationally invariant free fermions are described by the Hamiltonian
\begin{equation}
\label{eq_Hfree}
    \hat{H}_\text{free}=-\sum_{i=1}^{L} (\hat{c}_{i}^\dagger \hat{c}_{i+1}+\hat{c}_{i+1}^\dagger \hat{c}_{i}),
\end{equation} 
which has a doubly degenerated single-particle spectrum, except for two energies at the spectral edges. In the original calculation of $s^{(i)}_\text{free}$, see Ref.~\cite{Vidmar_2017}, the single-particle energy eigenstates were selected as
\begin{equation}
    |k\rangle=\frac{1}{\sqrt{L}}\sum_{i=1}^{L}e^{\text{i}\frac{2\pi}{L} ki}|i\rangle\;,
\end{equation}
with quasimomenta $k=-L/2+1,...,L/2$ and energies $\epsilon=-2\cos(\frac{2\pi}{L} k)$. We here argue that $s^{(i)}_\text{free}$ depends on the choice of single-particle energy eigenstates. For this reason, we redefine the latter as the following linear combinations:
\begin{equation}
    |\epsilon,+\rangle = \sqrt{1-\mu^2}|-k\rangle + \mu|k\rangle\;,
\end{equation}
and
\begin{equation}
    |\epsilon,-\rangle = \mu|-k\rangle - \sqrt{1-\mu^2}|k\rangle\;,
\end{equation}
for $k\ge 0$, except for $k=0$ and $L/2$, each corresponding to an undegenerated energy. The parameter $\mu$ describes the coupling between the quasimomenta of opposite signs. We assume $\mu$ is real and, thus, it belongs to the interval $[0,1)$. We plot $s^{(i)}_\text{free}$ against $\mu$ in Fig.~\ref{fig_e2}. Clearly, the volume-law coefficient $s^{(i)}_\text{free}$ varies with the coupling $\mu$, so that $s^{(i)}_\text{free}\approx0.5378$ is acquired only for $\mu^2=0$ and $0.5$. We note that when the OBC are imposed on the Hamiltonian form Eq.~(\ref{eq_Hfree}), the double degeneracy is lifted, and the single-particle energy eigenstates correspond to $|\epsilon,\pm\rangle$ with $\mu^2=0.5$. Remarkably, when we average the volume-law coefficient of the translationally invariant free fermions over all couplings, we arrive at $\langle s^{(i)}_\text{free}\rangle_\mu\approx 0.5428$. The latter is in almost perfect agreement with the volume-law coefficient in the delocalized regime of the Anderson model with PBC, $s^{(i)}\approx 0.5429$, see the inset of Fig.~\ref{fig_e1}(c).

\subsection{Bipartitions in quasimomentum space} \label{sec:bipartitions_momentum}

It is interesting that although the spectral statistics of the Anderson model agree with the GOE prediction at $\tilde{W}_\text{GOE}$ [for both boundary conditions, see Fig.~\ref{fig_1}(a) and~\ref{fig_speca}(b)], the entanglement entropy studied in Sec.~\ref{sec:bipartitions_real} does not. 
Since the study of the structure of eigenstates in Sec.~\ref{sec:structure_eigenstates} suggested the presence of nontrivial features not only in position space but also in quasimomentum space, we here study the entanglement entropies for bipartition in quasimomentum space, $\langle\langle S^{(k)}\rangle_m\rangle_H$. The latter is calculated from the eigenvalues of the following one-body correlation matrix,
\begin{equation}
\label{eq_J}
    J^{(k)}_{pq}=2\langle m| \hat{f}_{p}^\dagger \hat{f}_{q} |m\rangle-\delta_{pq},
\end{equation}
where $\hat{f}_{q}=\frac{1}{\sqrt{L}}\sum_{j=1}^{L} e^{\text{i}\frac{2\pi}{L}qj}\hat{c}_j$ annihilates a spinless fermion with a quasimomentum $q$. Therefore, the bipartition is performed in the quasimomentum space and the subsystem A is composed of $L_A$ contiguous quasimomenta. As previously, we fix the subsystem fraction to $f=\frac{L_A}{L}=\frac{1}{2}$ and define the volume-law coefficient as
\begin{equation}
s^{(k)}= \frac{\langle\langle S^{(k)}\rangle_m\rangle_H}{\frac{L}{2} \ln(2)}.
\end{equation}
We plot the volume-law coefficients $s^{(k)}$ as functions of the scaled parameters in the Wannier-Stark model and the Anderson model with OBC and PBC in Fig.~\ref{fig_e3}(a), \ref{fig_e3}(b), and \ref{fig_1}(b), respectively. We first note that the results for $s^{(k)}$ match $s^{(i)}_\text{free}$ in the limit of large $\tilde{F}$ and $\tilde{W}$, when the single-particle energy eigenstates are completely delocalized in the quasimomentum space but not the position space. The drift of the right shoulders of the curves from Figs.~\ref{fig_e3}(b) and \ref{fig_1}(b) can be explained by the vanishing of this regime in the nonstandard thermodynamic limit of the Anderson model. As for the variance $\sigma^2$ from Fig.~\ref{fig_ipr3}, the drift disappears when $s^{(k)}$ are plotted against $W$ (not shown). 

At first, it may seem surprising that $s^{(k)}$ do not vanish when the single-particle energy eigenstates are completely localized in the quasimomentum space. However, this can be easily explained for the Wannier-Stark and Anderson models with OBC, since the single-particle energy eigenstates have similar projections on single-particle states with opposite quasimomenta so that they become $|\epsilon,\pm\rangle$ with $\mu^2\rightarrow 0.5$ when $\tilde{F},\tilde{W}\rightarrow 0$. Therefore, each particle occupying $|\epsilon,\pm\rangle$ in a many-body energy eigenstate exists on both sides of the boundary separating the subsystem from the environment in the quasimomentum space and, so, it contributes to the entanglement entropy. Although this reasoning explains why the volume-law coefficient is close to $1/2$, it does not explain its perfect agreement with $s^{(i)}_\text{free}$, see the violet solid line in Figs.~\ref{fig_e3}(a) and~\ref{fig_e3}(b). Simultaneously, the single-particle energy eigenstates of the Anderson model with PBC become $|\epsilon,\pm\rangle$ with an arbitrary $\mu$ when $\tilde{W}\rightarrow 0$. This is reflected in the diminished value of the volume-law coefficient, $s^{(k)}\approx 0.5$, see the pink dashed line in Fig.~\ref{fig_1}(b).

Finally, the entanglement entropy in quasimomentum space of the Anderson model exhibits a peak in the vicinity of $\tilde{W}_\text{GOE}$ for both boundary conditions, see Figs.~\ref{fig_e3}(b) and \ref{fig_1}(b). The volume law coefficient $s^{(k)}$ near the critical point $\tilde{W}_\text{GOE}$ is lower but close to $s_\text{chaotic}^{(i)}$ and further increases with the system size $L$. We fit a second-order polynomial to the topmost part of the peak from Fig.~\ref{fig_e3}(b) to establish the maximal value of $s^{(k)}$. The difference
\begin{equation}
    \delta=s^{(i)}_\text{chaotic}-\text{max}[s^{(k)}]
\end{equation} 
closely follows the polynomial decay $aL^b$ with $b\approx -0.085$ and, thus, may vanish in the nonstandard thermodynamic limit, as illustrated in Fig.~\ref{fig_e3}(c). We highlight that the central assumption in the derivation of $s_\text{chaotic}^{(i)}$ is that the projections of single-particle energy eigenstates onto Wannier states, $\langle i | \alpha \rangle$, are normally distributed random numbers~\cite{PhysRevLett.125.180604}. It remains to be understood why this assumption seems to be valid in the Anderson model near $\tilde{W}_\text{GOE}$ in the quasimomentum space (i.e., for $\langle k | \alpha \rangle$) rather than in the position space (i.e., for $\langle i | \alpha \rangle$).

\section{Conclusions} \label{sec:conclusions}

In certain quadratic systems, such as the 1D Anderson model and 1D Wannier-Stark model, the crossover from spatially delocalized states (exhibiting localization in quasimomentum space and ballistic transport) to localization in position space takes place without any intermediate regime exhibiting quantum chaos and diffusive dynamics.
In the standard thermodynamic limit, all states are localized in the position space for any nonzero on-site potentials.
Yet we showed that in the nonstandard thermodynamic limit, i.e., upon proper rescaling of the model parameter with the system size, one can detect a well-defined transition point that signals the deviation from complete delocalization in the position space.

We argued that the transition point in the 1D Anderson model exhibits some unexpected properties.
In particular, it appears to be universal and nonuniversal at the same time, giving rise to the Janus-type character.
Among the universal properties, we observed the level spacing ratio complying with the GOE prediction, and the eigenstate entanglement entropy for bipartition in quasimomentum space approaching the value of Haar-random Gaussian states.
In contrast, the eigenstate entanglement entropy for bipartition in position space departs from the value of Haar-random Gaussian states, and may also be distinct from the value of translationally invariant free fermions.
These observations suggest that, in contrast to transitions from chaos to localization, these transitions require special measures for their characterization, whose generality is yet to be understood.

\acknowledgements 
We acknowledge discussions with M.~Mierzejewski and M.~Hopjan. 
Numerical studies in this work have been partially carried out using resources provided by the~Wroclaw Centre for Networking and Supercomputing, Grant No. 579 (P.{\L}., M.L.).
L.V. acknowledges support from the Slovenian Research and Innovation Agency (ARIS), Research core funding Grants No.~P1-0044, No. N1-0273, No. J1-50005, and No. N1-0369.

\appendix

\section{Density of states in Wannier-Stark and Anderson models with OBC} \label{Appendix_DOS}

\begin{figure}[!t]
\includegraphics[width=\columnwidth]{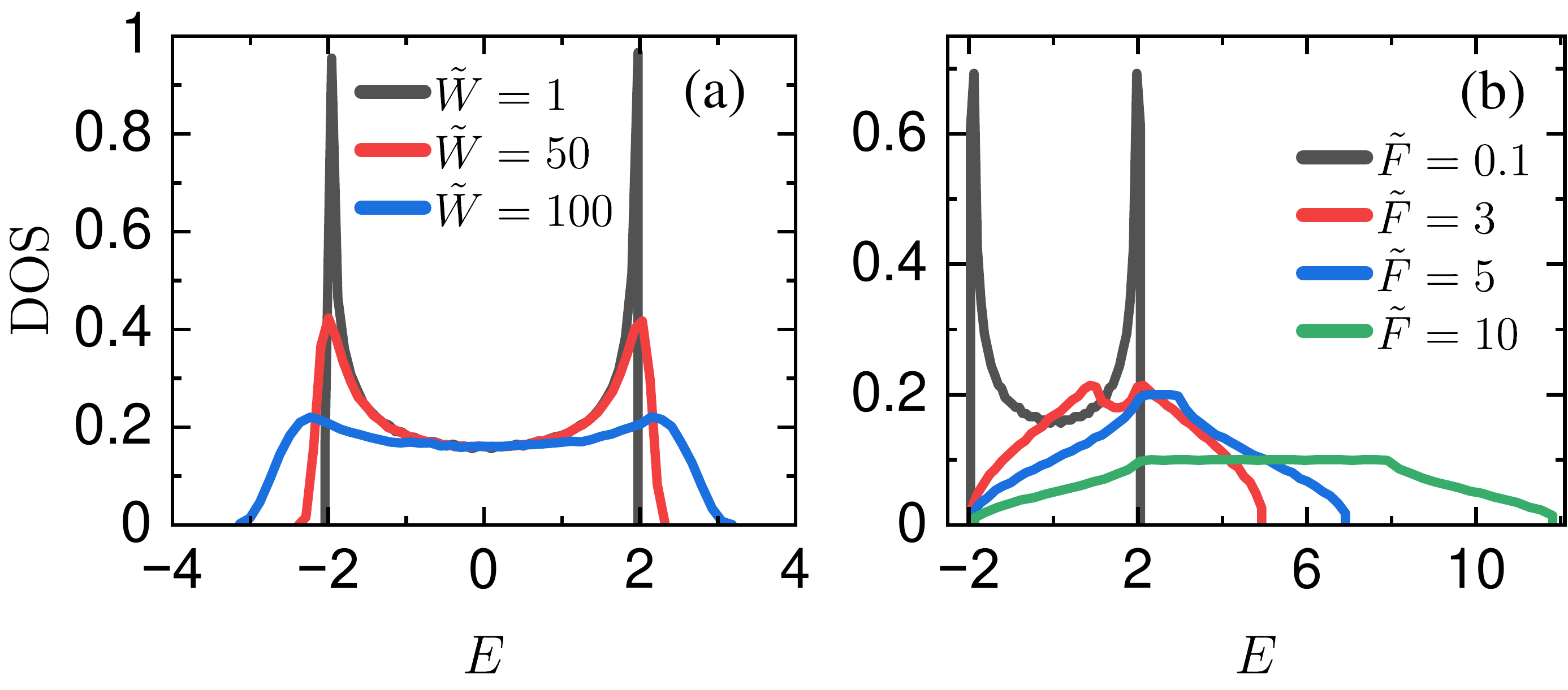}
\caption{The density of states in (a)~Anderson model and (b)~Wannier-Stark model with OBC. We consider different values of scaled on-site potentials, $\tilde{W}$ and $\tilde{F}$, and focus on the system size $L=2500$. Numerical calculations were averaged over $100$ disorder realizations in (a).}
\label{fig_dos}
\end{figure}

The density of states for the 1D Anderson model with OBC is shown in Fig.~\ref{fig_dos}(a), while that for the 1D Wannier-Stark model with OBC is shown in Fig.~\ref{fig_dos}(b). For the 1D Anderson model, all numerical results are averaged over 100 Hamiltonian realizations. We consider different values of the scaled on-site potentials, $\tilde{W}$ and $\tilde{F}$, and focus the system size $L = 2500$. However, we have verified that the properties of the density of states, both qualitative and quantitative, are independent of $L$. Although deep in the delocalized regime the density of states exhibits two peaks near the edges of the spectrum, these peaks disappear and the density of states becomes featureless as the change in the scaled on-site potential, $\tilde{W}$ or $\tilde{F}$, drives the system through the eigenstate transition.

\section{Spectral statistics in Wannier-Stark and Anderson models with OBC} \label{Appendix_spectral}

\begin{figure}[!h]
\includegraphics[width=\columnwidth]{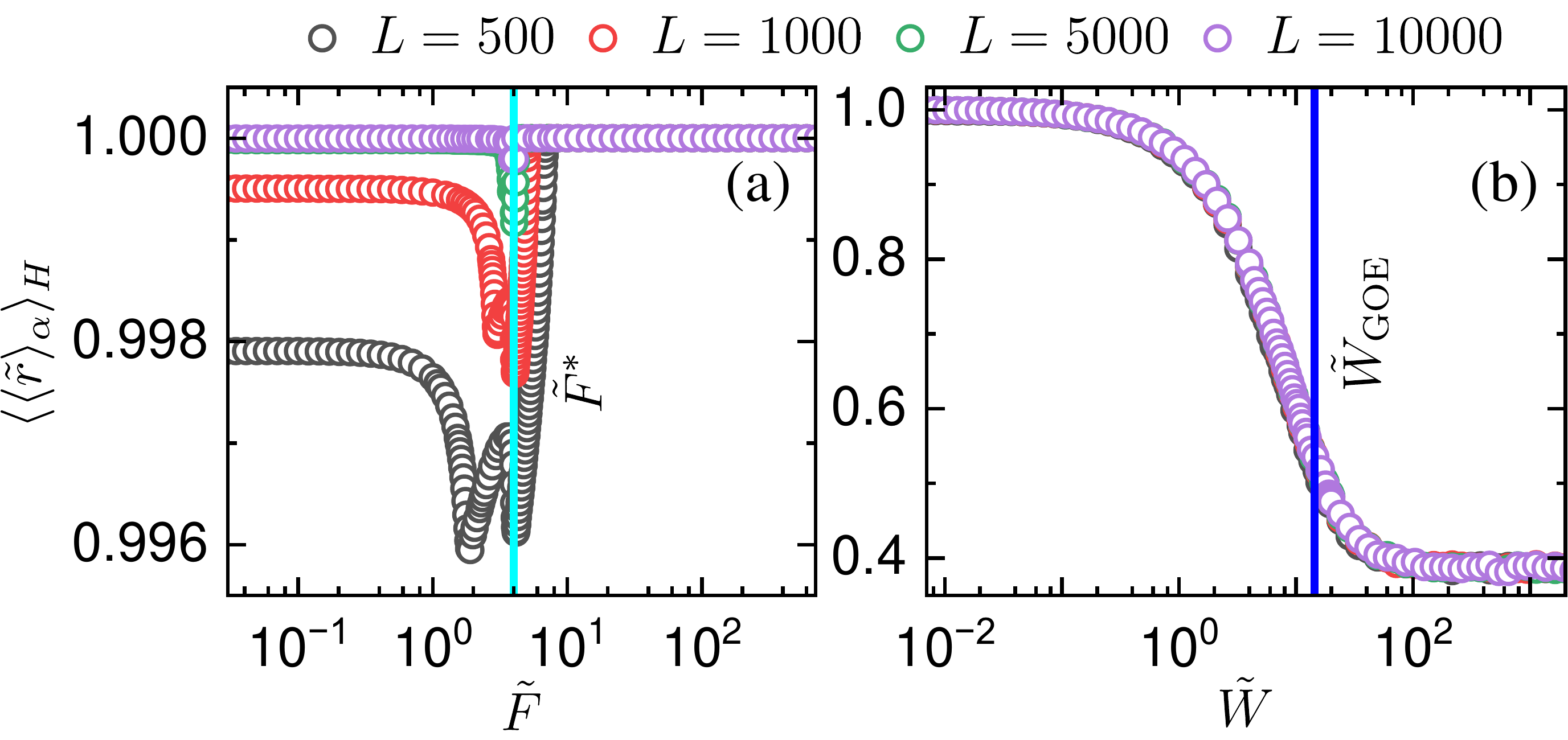}
\caption{The level spacing ratio in (a) Wannier-Stark model and (b) Anderson model with OBC. We consider $L=500, 1000, 5000$ and $10000$. The cyan vertical line in (a) marks $F^*$, while the blue vertical line in (b) marks $\tilde{W}_\text{GOE}$. Numerical calculations were averaged over $100$ single-particle energy eigenstates in the middle of energy spectrum. They were additionally averaged over $100$ disorder realizations in (b). Legend provided above the figure is valid for all panels.}
\label{fig_speca}
\end{figure}

The spectral statistics of the Wannier-Stark model are provided in Fig.~\ref{fig_speca}(a). In the nonstandard thermodynamic limit, the mean ratio $\langle\langle\tilde{r}\rangle_\alpha\rangle_H=1$ in the whole range of scaled fields $\tilde{F}$, as expected for a ladder structure. However, for finite system sizes, it develops a shallow minimum near the eigenstate transition at $\tilde{F}^*$, indicated by a vertical cyan line.

The spectral statistics of the Anderson model with OBC are illustrated in Fig.~\ref{fig_speca}(b). The mean ratio $\langle\langle\tilde{r}\rangle_\alpha\rangle_H$ is a smooth function of the scaled disorder strength $\tilde{W}$. It interpolates between $\langle\langle\tilde{r}\rangle_\alpha\rangle_H\approx 1$ at small $\tilde{W}$, characteristic of equidistant single-particle energy levels, and $\langle\langle\tilde{r}\rangle_\alpha\rangle_H\approx \tilde{r}_\text{P}$ at large $\tilde{W}$, indicating an uncorrelated single-particle spectrum with Poisson statistics. Consequently, it crosses the value predicted for the GOE ensemble, i.e., $\tilde{r}_\text{GOE}$, at $\tilde{W}_\text{GOE}\approx 14.36$ (see also Refs.~\cite{PhysRevE.86.011142,PhysRevE.100.022142}). The latter value was established from the linear regression of $\langle\langle\tilde{r}\rangle_\alpha\rangle_H$ performed in the interval $\tilde{W}\in[3,20]$ and it is marked by a vertical blue line.

\section{Critical scaled field strength in Wannier-Stark model} \label{Appendix_ipr}

\begin{figure}[!b]
\includegraphics[width=\columnwidth]{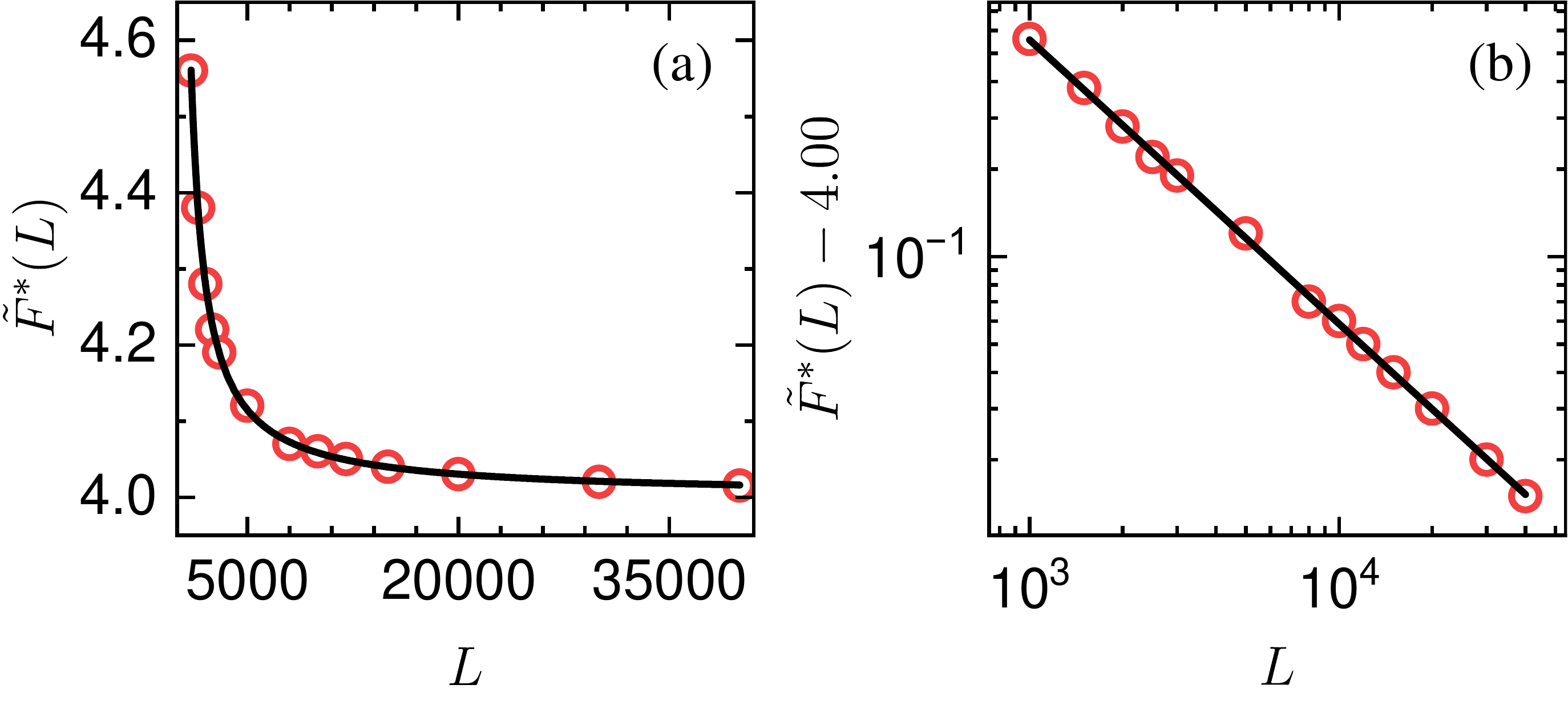}
\caption{(a) The critical scaled field strength $\tilde{F}^*$ plotted as a function of the system size $L$ in the Wannier-Stark model. (b) $\tilde{F}^*-4.002$ plotted against $L$. The red points are results of numerical calculations, while the black line is the least-squares fit of (a)~$y=a x^{b}+c$ with $a\approx540.75,\;b\approx-0.10,\;c\approx4.00$ and (b)~$y=ax+b$ with $a\approx-0.98,\;b\approx2.73$.}
\label{fig_ipra}
\end{figure}

In this section, we numerically determine the critical scaled field strength in the Wannier-Stark model using the inverse participation ratio in the quasimomentum space. Specifically, we establish the position of a sudden jump in $\text{ipr}^{(k)}$ for each considered system size $L$, denoting it as $\tilde{F}^*(L)$. The latter is defined as the point at which $\text{ipr}^{(k)}$ exceeds its value within the plateau, exhibited in the regime of cD in the quasimomentum space, by more than one standard deviation as the scaled potential is decreased. As shown in Fig.~\ref{fig_ipra}(a), the critical field strength $\tilde{F}^*(L)$ appears to scale polynomially with the system size with the points obtained from numerical calculations, closely following the least-squares fit $\tilde{F}^*(L)=4.00+a L^{b}$, where $a\approx540.75$ and $b\approx-0.10$. This is further validated in Fig.~\ref{fig_ipra}(b), where we plot $F^*(L)-4.00$ and confirm its linearity on a log-log scale.

\FloatBarrier
\bibliographystyle{biblev1}
\bibliography{references}

\end{document}